**Following the Herd: The Influence of the Bandwagon Heuristic on E-Commerce Shoppers**

Min Xiao
Associate Professor
Wichita State University
Min.xiao@wichita.edu
316-978-6051
ORCiD: 0000-0001-5367-4322
Elliott School of Communication, Wichita State University
1845 Fairmount Street, Campus Box 031, Wichita, KS, USA 67260

Paul Myers
University of Missouri
pcmnm9@missouri.edu

**Abstract**

E-commerce has grown over the years, and the COVID-19 pandemic has further invigorated consumers' enthusiasm to shop online. To consumers, product reviews and related metrics are useful resources to help them make informed purchase decisions. This research study examined how product review scores, the number of reviews, product involvement (i.e., product types), and product review valence affect consumers' shopping decisions. Specifically, two online experiments were conducted to examine how product review scores interact with the number of reviews, product involvement, and review valence in affecting consumers' bandwagon perception and purchase intention. A mock e-commerce site that displays products and product review metrics was constructed for the purpose of the research. Experiment participants were recruited from the Amazon's Mechanical Turk (MTurk) platform. The findings suggest that product review scores, and the scores' interaction with product involvement and review valence, influenced consumers' bandwagon perception. Moreover, the bandwagon perception mediated the correlation between product review scores and product purchase intention.

E-commerce has expanded rapidly to the point that most modern consumers are highly aware of it and have actively purchased items through an online retailer. Over 200 million U.S. consumers have shopped online, and this number is likely to grow (eMarketer, 2017). The rapid advancement of media technology brings consumers an array of resources that help them make purchase decisions. Oftentimes these consumers look to user-generated electronic word-of-mouth (eWOM), in the form of reviews and product review scores, to reduce their concerns and uncertainty about the product they are considering purchasing (Chen & Xie, 2008). These reviews and review scores are meant to serve as a measure of a product's quality and value. While reviews and review scores are intertwined, the terms refer to different portions of the product evaluation. The term review refers to the text-based or sentiment portion of the content, while the term review score refers to the numerical value assigned by the reviewer (Hu et al., 2014).

The advancement of media technologies has created an overabundance of information offerings that complicates the process of information evaluation (Fogg et al., 2002). Individuals are not always motivated to process information extensively, and they often attempt to evaluate information in an efficient manner that conserves mental energy (Fiske & Taylor, 1991). Thus, individuals often utilize heuristics (i.e., cognitive shortcuts) to facilitate the processing of information. As the number of product choices for e-commerce is ever-growing, consumers look for ways to simplify the decision-making process. Product reviews and related metrics are often utilized by consumers to reduce effort and search time when evaluating items within a given product category.

The current research focused on studying the bandwagon heuristic. This heuristic has been investigated because of its innate association with product review metrics. With the bandwagon heuristic as the centerpiece of the investigation, the current research examined how product review score interacted with number of reviews, product involvement, and review valence in affecting the way consumers evaluated product-related information. The significance of the study can be discussed from the following perspectives.

This study is an in-depth analysis on a heuristic closely related to social influence in the context of e-commerce. Product review scores are some of the most important cues that shape consumers' impression of a product. Though the direct impact of product review metrics on the use of bandwagon heuristic has been examined (e.g., Xu, 2013; Waddell, 2018), the indirect or interaction effects between the metrics and the psychological state can be studied in more depth.

Many studies have directly correlated the information cues with information processing outcomes without measuring the specific heuristics employed by study participants (e.g., Wood et al., 1985; Wu & Lin, 2017). However, a psychological state, such as a heuristic, may exist between the cue and information processing outcomes that mediates the correlation between the two (Bellur & Sundar, 2014). If researchers do not examine or measure the psychological state, the linkage established between cues and information processing outcomes might be spurious, and the study conclusion might be constructed based on a questionable foundation. Hence, this study was designed to address this limitation identified in extant studies by measuring the bandwagon heuristic and analyzing the mediation effects associated with the construct.

Other than mediation effects, the current research examined the interaction effects associated with the involvement construct. Wang et al. (2023) conducted a meta-analysis on empirical studies that examined the bandwagon heuristic and its influence on information credibility assessment. The meta-analysis listed an array of factors cueing the use of the bandwagon heuristic. However, the involvement construct is a notable variable missing from the list. Consumers exert differing levels of effort while assessing product-related information, depending on the value or relevance of the product to the consumers' intention. An important product may motivate consumers to process the information extensively, but a trivial product is unlikely to motivate a consumer in the same way. Such a difference in products' level of importance can be conceptually associated with the difference in product involvement levels. As product involvement levels vary, the influence of information cues on consumers' usage of heuristics changes.

Would the influence of bandwagon heuristic and related information cues on behavioral intention vary as the level of product involvement changes? Finding an answer to this question is both theoretically and managerially meaningful. The extent to which a heuristic is used depends on an individual's level of involvement with an issue or entity (Chaiken, 1980). A study examining the impact of product involvement on the usage of the bandwagon heuristic offers significant theoretical contributions to our understandings about heuristic information processing in the context of online shopping. Moreover, the findings of the research would validate the discoveries of existing studies and expand the boundary of the knowledge regarding psychological mechanisms of how information cues activate a heuristic that influences behavioral intention. Managerially, an answer to the question would offer strategic recommendations to marketers in terms of how they could utilize product review metrics on an e-commerce site to optimize their eWOM marketing efforts.

Two online experiments were conducted on Amazon Mechanical Turk (MTurk). Mock webpages were created as the experiment stimuli. The mock webpages manipulated product review scores, number of reviews, product involvement, and review valence. The investigation started with an inquiry about how product review scores, number of reviews, and product involvement interacted and influenced the bandwagon heuristic. Study 2 examined how product review scores, product review valence, and product involvement affected the bandwagon heuristic. The mediating impact of the bandwagon heuristic on the correlation between product review scores and purchase intention was examined in both studies.

## Literature Review

The theoretical foundation of the current research lies in the literature of the bandwagon heuristic. Over the years, scholars have studied the bandwagon heuristic under various contexts such as political elections (e.g., Schmitt-Beck, 2015), economics (e.g., Leibenstein, 1950), and communication (e.g., Walther & Jang, 2012; Lee & Sundar, 2013). In the past decade, the bandwagon heuristic examined in the studies of mass communication has been associated with the theoretical framework of the MAIN model (Sundar, 2008). In studies that employed the MAIN model (e.g., Li & Sundar, 2022; Luo et al., 2022), various types of information cues associated with the bandwagon heuristic, as well as the accompanied information processing outcomes, have been discussed and documented. As the name of the construct indicates, the bandwagon heuristic is a form of cognitive heuristics. The goal of using heuristics is to expedite the speed of information processing (Chaiken et al., 1989). Other than the MAIN model, the concept of psychological heuristic has been discussed by many scholars who study dual process models such as the Elaboration Likelihood Model (ELM) and Heuristic-Systematic Model (HSM). To give fellow scholars a comprehensive view of the bandwagon heuristic and its potential impact on information processing outcomes, existing literature related to the dual process models has been reviewed in the next section preceding the discussion of the MAIN Model.

**Dual Process Models**

Psychological theories or models about information processing, such as the HSM and the ELM, claim that two routes or modes exist when humans process information (Chaiken, 1980; Petty & Cacioppo, 1990). The first mode is fast and intuitive since the information is processed through heuristics, which are formed based on past experiences or generalized opinions of others (Evans, 2003; Petty et al., 2015). When using the first mode an individual exerts minimum cognitive effort in processing a message. In contrast, the second mode demands more cognitive deliberations. The second mode, often called central or systematic processing, is used by an individual when the person is motivated and cognitively capable of processing the information (Chaiken & Maheswaran, 1994). Humans have limited cognitive capacity, meaning that we tend to use the least amount of cognitive effort to make quickest judgments (Lang, 2000). Thus, the first system, the heuristic mode, is often employed in numerous situations. The tendency to use heuristics is so strong that even when a person is motivated and capable of processing information, heuristics are still used to some extent (Petty & Brinol, 2012).

In dual process models, such as the HSM and ELM, information cues and psychological heuristics are closely associated constructs. Many studies have examined the impact of information cues on the use of heuristics and the subsequent information processing outcomes. For instance, individuals are likely to rely on cues, such as the length of messages (Wood et al., 1985), number of arguments (Petty & Cacioppo, 1984), and attractiveness of communicators (Eagly & Chaiken, 1975), to help them evaluate the persuasiveness of a message. However, a significant weakness of the mentioned studies is that the specific type of heuristic employed by research participants was, unfortunately, not directly measured by the researchers (Eagly & Chaiken, 1993). Hence, assumptions or findings reported by the studies were concluded based on indirect or circumstantial evidence. In fact, even a few contemporary studies, in which heuristics and information processing outcomes were examined (e.g., Wu & Lin, 2017; Xiao & Myers, 2022), have a similar issue. That is, the specific type of heuristic used by study participants was not directly measured or operationalized. Thus, it is unclear if the examined cues (e.g., length of messages) stimulated study participants to use the heuristic that the researchers intended to examine. To address such a limitation, the current research directly measured the heuristic so that solid empirical evidence could be obtained to help scholars analyze the impact of the bandwagon heuristic.

In contrast, some other empirical studies, typically the ones that employed the MAIN model, have measured the specific types of heuristics. In those studies, the information cues were often manipulated in different experimental conditions, and the association between the cues and bandwagon heuristic was analyzed statistically (Bellur & Sundar, 2014). For instance, scholars have discovered that cues, such as the number of product reviews (Sundar et al., 2008), retweets (Waddell & Sundar, 2017), and views of an article (Xu, 2013) activate the use of the bandwagon heuristic.

**The MAIN Model**

The letters in the MAIN model denote Modality (M), Agency (A), Interactivity (I), and Navigability (N). The model aims to explain how technological affordances affect the way an individual processes information presented on digital media platforms (Sundar, 2008). The common thread among the HSM, ELM, and MAIN model is that all three theoretical frameworks discuss the impact of heuristics. The difference is that dual process models address the importance of both heuristic and systematic processing, while the MAIN model focuses on examining the association between information cues and the use of heuristics.

Sundar (2008) defined heuristic as a mental generalization created based on acquired knowledge or experience. According to the MAIN model, a technological affordance or an action possibility can serve as an information cue (Evans et al. 2017; Sundar 2008). Cues, though not heuristics per se, contain information or elements associated with heuristics and hold the potential to trigger heuristics (Sundar, 2008). In other words, information cues are working in tandem with heuristics in affecting information processing outcomes.

Information cues and heuristics are two related yet different constructs. One should note that the successful manipulation of a certain cue may not always induce the use of a heuristic that researchers intend to examine. Thus, researchers must measure the heuristic to ensure that the manipulation of cues successfully triggers the heuristic of interest. One of the heuristics that are relevant to the theme of the current research is the bandwagon heuristic. Elements on an e-commerce site, such as product review scores, can be considered information cues that trigger the bandwagon heuristic.

**The Bandwagon Heuristic**

The bandwagon heuristic, specifically situated in the context of the current research, represents a herd mentality (e.g., the product is of high quality if others consider the product this way). Existing research has operationalized the bandwagon heuristic as self-reported bandwagon perception (Xu, 2013; Waddell & Sundar, 2017). Thus, the current research followed the procedures employed by the previous studies and measured the bandwagon heuristic as the bandwagon perception.

Product review metrics, such as product review scores and the number of reviews, are product quality or popularity indicators. In the MAIN model, these indicators are often conceptualized as bandwagon cues. Sundar et al. (2017) defined bandwagon cues as indicators embedded in the user interface that reveal system-generated information about other users' opinion of a certain entity.

Product review metrics, conceptually regarded as bandwagon cues, are potential triggers of the bandwagon heuristic. Many empirical studies that examined the bandwagon heuristic have unearthed a positive correlation among bandwagon cues, the heuristic, and information processing outcomes. Sundar et al. (2008) found that the increase in product review scores led to the increase in the bandwagon perception, which then, influenced an individual's purchase intention. In a similar study, Sundar and colleagues (2009) discovered that bandwagon cues, a combination of product star ratings and number of reviews, positively affected the bandwagon perception and purchase intention of a product. However, the study did not delineate the differences between review scores and number of reviews in terms of their psychological impact on consumers. Xu (2013) centered the investigation on the bandwagon perception and found that number of votes on a news article significantly influenced the bandwagon perception, perceived news credibility, and click rate.

Some empirical findings are inconsistent with what have been reported thus far. For instance, Bode et al., (2021) did not discover a significant bandwagon effect induced by the bandwagon cues, while Na (2015) found a negative correlation between the level of bandwagon support and the information processing outcomes. The inconsistency in empirical findings can be explained by many factors (e.g., different dependent variables, different study contexts, etc.). Regardless of the causes, such an inconsistency is a call for scholars to continue the research of the bandwagon heuristic so that more evidence in explaining the phenomenon can be offered in a consistent manner. The following hypothesis was proposed to guide this investigation.

**H1**. Product review scores positively influence the bandwagon perception.

The influence of the number of reviews has been documented in existing studies. Flanagin and Metzger (2013) uncovered a positive relationship between the quantity of movie reviews and the perceived trustworthiness of reviews and review scores. Fu and Sim (2011) unearthed a statistically significant association between the existing viewership count of a YouTube video and the future viewership count of the video. However, the goal of many studies reviewed thus far, either in this section or in the previous section of dual process models, was not about investigating the influence of heuristics. In those studies, the heuristic used by study participants was not directly measured, and it was unclear if the increase in the number of reviews or comments affected consumer perception via the influence of the bandwagon heuristic. A research study is needed to ascertain the influence of the number of reviews on the bandwagon heuristic.

A goal of the current research is to investigate how bandwagon perception is affected by the product review score's interaction with the number of reviews. The increase in the number of reviews is likely to amplify the impact of review scores on the bandwagon perception because such an increase represents an expansion of a product's fanbase. Hence, the following hypothesis was proposed.

> **H2**. The influence of product review scores on the bandwagon perception is moderated by the number of reviews.

More importantly, heuristics, such as the bandwagon heuristic, need to be regarded and analyzed as mediators between cues and information processing outcomes (Sundar et al., 2008; Sundar et al., 2009; Waddell & Sundar, 2017). Bellur and Sundar (2014) reviewed a series of academic research (e.g., Chen & Chaiken, 1999; Kahneman & Tversky, 1973) that investigated heuristics' impact on individuals. The researchers concluded that heuristics should be treated as intervening psychological states (i.e., mediators) that help facilitate the decision-making processes (Bellur & Sundar, 2014). Information cues (e.g., product review scores) serve as the trigger for a heuristic (e.g., the bandwagon heuristic), and the cues contain information related to a heuristic (Sundar et al., 2017). In other words, an information cue reminds an individual about a generalized opinion formed based on experience (e.g., high score means high product quality), and such an opinion may further affect information processing outcomes (e.g., I should buy the product).

Previous research studies in marketing have examined how product review scores influenced consumer purchase decisions. Hu and colleagues (2014) investigated 4,000 book review ratings on Amazon.com. The researchers discovered that book review scores indirectly influenced sales via review sentiment, and the recency of reviews significantly influenced sales. Moe and Trusov (2011) analyzed product review data of a national retailer and found a significant direct impact of existing product review scores on subsequent review scores and an indirect impact of existing product review scores on sales. However, the discovered effects were not long-lasting since the introduction of other factors in the model rendered the impacts ineffective. Dellarocas et al. (2007) found that the addition of movie review scores in a statistical model, including factors, such as prerelease marketing investment and theater availability, significantly improved the predictive accuracy of the model. In the context of communication and media studies, several researchers have examined the bandwagon perception as a mediator in their works. For examples, bandwagon perception has been examined as a mediator between the product review score and purchase intention of a camera (Sundar et al., 2008), between the valence of social media posts and television program enjoyment (Waddell & Sundar, 2017), and between the number of likes and perceived newsworthiness of an article (Waddell, 2018). The empirical evidence unearthed by existing studies suggest that the bandwagon perception is likely to be the missing piece that mediates the relationship between product review score and purchase intention in the context of the current research. The following hypothesis was proposed.

**H3.** The bandwagon perception mediates the correlation between product review score and purchase intention.

**The Bandwagon Heuristic and Product Involvement**

In addition to the mediation effect, the varying intensity of the bandwagon heuristic and the associated factors should be examined. The impact of the bandwagon heuristic has been examined from various angles. Much of the literature has adhered to the principles proposed in the MAIN model. Therefore, the studies may have overwhelmingly examined the bandwagon effect from a heuristic or peripheral processing perspective, whereas the potential impact exerted by factors relevant to effortful processing (e.g., involvement) may have been overlooked. According to the dual process models, heuristics are ubiquitously used even when an individual processes information effortfully (Chaiken & Maheswaran, 1994; Petty & Cacioppo, 1984). The impact of information cues on the use of heuristics may vary as the importance of an issue, or one's level of involvement with an issue, changes. In other words, the degree of effort one exerts in processing a piece of information may alter how the bandwagon cues affect the use of the heuristic as well as the information processing outcomes.

Extant research related to the bandwagon heuristic has examined an array of moderators. The moderators include the operational form of cues (Jin & Phua, 2014), cue levels (Westerman et al., 2012), message valence (Hayes et al., 2018), message modality (Sundar et al., 2021), and cultural differences (Hofstede et al., 2010). However, a notable variable missing from the list is involvement. In fact, the involvement factor might have not been examined extensively in the studies that employed the MAIN model (e.g., Hayes et al., 2018; Westerman et al., 2012).

Depending on the level of involvement, the same cue can trigger a heuristic in different levels of intensity. For instance, the length of a news article can elicit a "length is strength" heuristic in low involvement conditions, while a "length is prominence" heuristic can be elicited in high involvement conditions (Bellur & Sundar, 2014). Another example can be observed by reviewing the study conducted by Beltramini and Stafford (1993). The researchers found that a product label was regarded as a peripheral information cue by study participants in low product involvement conditions, but the same label functioned as a central factor in high product involvement conditions (Beltramini & Stafford, 1993). Thus, one can deduce based on the empirical evidence that the influence of product review metrics on the use of the bandwagon heuristic and information processing outcomes (e.g., purchase intention) changes as the level of involvement varies. These examples demonstrate the importance of examining involvement, a concept closely associated with dual process models and the MAIN model, as a potential moderator that affects the intensity of the bandwagon effect.

The construct of involvement has been expounded in literature of social and consumer psychology that employed dual process models. According to Petty and colleagues (1983) the construct represents the degree of relevance, consequence, or connection of an issue to one's personal life. High involvement issues usually encourage individuals to process information systematically or effortfully, whereas low involvement issues may not motivate individuals the same way (Eagly & Chaiken, 1983). As discussed earlier in the manuscript, extant studies about dual process models reveal that even when an individual processes information effortfully, the individual is still likely to use information cues and the associated heuristics to help the person process information expeditiously (Chaiken, 1980; Petty & Brinol, 2012).

Involvement, situated in the e-commerce context of the current research, is conceptualized and operationalized as product involvement. Though differences exist, product involvement is similar to the construct of involvement explained in dual process models. A household product (e.g., napkins) offers simple functions (e.g., cleaning) at a low price that may lead consumers to consider the product trivial. In contrast, an expensive electronic good (e.g., computer) offers sophisticated functions (e.g., graphic design) that many consumers may consider the product important. In other words, products can be distinguished based on their degree of relevance, consequence, or connection to one's personal life (Petty et al., 1983; Zaichkowsky, 1985). Inexpensive, fast-moving consumer goods are usually considered low involvement products due to their low value to one's life, while costly and technologically sophisticated consumer electronic goods are usually considered high involvement products because of their potential in improving one's work or life quality (Zaichkowsky, 1985).

Extant studies that have examined the bandwagon perception and product reviews tend to use just one type of products such as books (Hu et al., 2014), movies (Flanagin & Metzger, 2013), nutritional supplements (Hong & Pittman, 2020), or digital cameras (Sundar et al., 2009). The difference between product categories can be conceptually explained by differences in product involvement levels. In other words, products examined by the reviewed studies were predominantly within the same product category or within the same product involvement level, and the moderating impact of product involvement was overlooked.

Based on the literature reviewed thus far, one may infer that the influence of product review scores on the bandwagon perception is likely to be stronger when the product involvement level is high than when it is low. To many consumers, a high involvement product is usually the one that is more valuable or useful than a low involvement product. As a result, consumers tend to be more careful when they are evaluating information related to a valuable product than when they are evaluating information related to a trivial product. Thus, the following hypothesis was proposed.

**H4.** The influence of product review scores on the bandwagon perception is moderated by product involvement levels.

**The Impact of Product Review Valence on the Bandwagon Heuristic**

In addition to product review scores and product involvement, the valence of review is likely to affect consumers' view of a product. Previous studies have documented the power of review valence in affecting consumer decisions from various angles. Zhang and colleagues (2010) employed regulatory focus theory and investigated the association between review valence and perceived persuasiveness of reviews. They discovered that individuals with a promotion goal perceive positive reviews as more persuasive than negative ones, whereas individuals with a prevention goal perceive negative reviews as more persuasive. Lee and Youn (2009) investigated how product review valence affects consumers on different media platforms. The researchers found that consumers are more willing to recommend a product to others when they encounter positive reviews on a brand's website than when they read positive reviews on personal blogs, and consumers are not likely to recommend any product when they read negative reviews posted on any platforms. Hong and Pittman (2020) studied how product review scores, the number of reviews, and review valence (all negative or all positive) affected perceived credibility of reviews. They found that positive reviews led experiment participants to trust star ranking system, while negative reviews led participants to pay more attention to the number of reviews or the content of reviews than to star ranking system.

Similarly, academic literature examining the bandwagon heuristic has documented the significant influence of user-generated review/comment on the heuristic. Lee et al. (2020) found that positive comments accompanying a news article led their study participants to view the subject portrayed in a news article positively, while negative comments led participants to view the subject negatively. Waddell and Sundar (2017) uncovered an overpowering influence of negative tweets on television audiences' bandwagon perception and their perception of a program. In a related study, Waddell (2018) found that negative comments, relative to positive ones, significantly decreased the bandwagon perception and negatively affected perceived news article credibility. The overwhelming influence of negative tweets or comments on consumers can be attributed to the negativity bias, a tendency of individuals to pay more attention to negative events than to positive events (Ito et al., 1997; Rozin & Royzman, 2001).

The findings of existing studies offered scholars insights regarding product review valence and its influence on consumers. However, there are three limitations in the reviewed studies that can be addressed by the current research. First, research studies of product reviews have an array of foci, yet the bandwagon heuristic is not one of them (e.g., Maslowska et al., 2019; Zhao et al., 2020; etc.). Second, even when the empirical study has a focus on the bandwagon heuristic (e.g., Lee et al., 2020), the study was not contextualized in an advertising (or marketing) context. Third, many studies manipulated and operationalized the construct of review valence as a negative/positive dichotomy (e.g., Zhang et al., 2010); however, product reviews are usually mixed, meaning that both positive and negative reviews are presented on the webpage simultaneously to the potential consumers. Thus, the current study has been designed to address the identified limitations.

Compared to numerical product review scores, the content of product reviews offers consumers richer and more contextualized information about a product. Thus, many consumers may consider product reviews a more important source of product-related information than the review scores. However, it is time consuming for consumers to process the content of reviews extensively. Hence, consumers' impression of reviews is likely to be formed based on a rough summary of the content. Such an impression is manifested as one's perception of the sentiment or tone of reviews (i.e., review valence). If the general sentiment of reviews is positive, the impact of review score on bandwagon perception may be enhanced because consumers are likely to value the input offered in product reviews; however, if the sentiment of reviews is negative, the impact of review score on bandwagon perception may be mitigated for the same reason.

**H5.** Product review valence moderates the correlation between product review score and bandwagon perception.

## Study 1

**Method**

A 2 (product review scores: low vs. high score) × 2 (number of reviews: small vs. large number of reviews) × 2 (product involvement: low vs. high product involvement) between-subjects experiment was conducted on MTurk. The MTurk screening criteria was HIT Approval Rate ≥ 99%, Number of HIT Completed ≥ 1,000, Location is U.S., and Did not Participate in Pretest. Attention check questions (e.g., recall product type, review score level, etc.) were placed in the Qualtrics questionnaire. The participants who failed to provide a reasonable answer to the attention check questions and those who did not complete the questionnaire were not included in the data analysis. A total of 406 responses were analyzed. Among all participants, around 51% were female ($n = 207$) and 48% ($n = 195$) were male. The average age of participants was 40.62, while the average income was $70,436. Regarding the ethnicity, most participants were White ($n = 307$), and the rest were African American ($n = 43$), Asian ($n = 41$), and Hispanic/Latino ($n = 24$). Each participant was awarded $1 for their participation. The research project was approved by the IRB committee of the researchers' institution. All participants signed the informed consent electronically in the Qualtrics questionnaire.

*Pretest*

A pretest ($n = 75$) was conducted on MTurk to determine the products to be displayed on the mock website. Twenty different types of products were tested in the pretest (e.g., bubble bath, facial tissues, smartphone, laptop computer, automobiles, etc.), and the products examined in the pretest were selected based on stimuli examined in previous studies (e.g., Atkinson & Rosenthal, 2014). Product involvement was measured by a 20-item, 7-point scale developed by Zaichowsky (1985). This product involvement scale was also utilized as manipulation check questions in the main experiments. A product involvement index was calculated for each type of product measured in the pretest based on the instruction of Zaichowsky (1985). The results revealed that bubble bath ($M = 64.54$, $SD = 32.93$) and a laptop computer ($M = 122.11$, $SD = 17.43$) received the lowest and the highest scores; thus, these products were manipulated in the experiment as the low and high involvement products. Participants were awarded $0.5 for their participation in the pretest.

*Stimuli*

A mock site (see examples in Appendix) was created. The site contains eight different pages for different experimental conditions. The site was created exclusively for the experiment and the design and layout of the webpages mimicked the style of a typical e-commerce site. Generic pictures of the products (i.e., bubble bath and laptop) were posted on webpages. The pictures were acquired online. No brand-related information was displayed in any of the pictures. Fictional brand names were used and listed on the webpages. Specifications (e.g., price, CPU speed, fragrance of bubble bath, etc.) of products were listed on the page, and the information was adapted from existing e-commerce sites (e.g., Amazon, Target, BestBuy, etc.).

Product review scores were manipulated by using a 5-point, star-based rating system on the webpage. The high score was 4.9 stars, while the low score was 1.2. The large number of reviews was 572, while the small number of reviews was 5. The level of review scores and the number of reviews were adapted based on values used by existing studies (Hong & Pittman, 2020; Sundar et al., 2008).

*Procedure of the Experiment*

In the experiment, participants were asked to read a message stating that the purpose of the study was to help a brand better market its products online and that fictional brand names were utilized to protect the identity of the brand. After reading the message, each participant was randomly assigned a different webpage URL that would direct each participant to an experiment condition. Participants were advised to visit the site on their computer for the best viewing experience. Participants were informed to return to the Qualtrics questionnaire after their visit to the site.

*Measurement*

In addition to the manipulated variables, the bandwagon perception was measured by an instrument adapted from Xu (2013). The items, "How likely are other people to like this product," "How likely is it that other people would think this product is valuable," "How likely are other people to recommend this product to their friends," "How likely is it that other people would share the information about this product with their friends," "How likely is it that other people would rate this product positively," were measured on a 7-point scale ranging from (1) Not at all likely to (7) Highly likely. All five items used by Xu (2013) were adapted and utilized in the current research. The measurement scale for the bandwagon perception exhibited excellent reliability ($\alpha$ = .960). Manipulation check questions (e.g., the review score that you saw on the webpage was (1) very low… (7) very high) were placed in the questionnaire of the main experiment. Purchase intention ($\alpha$ = .969) was measured on a 7-point Likert scale (i.e., "I would consider buying this product," "It is possible that I would buy this product," "I will purchase this product when I need it," "I intend to buy this product.") adapted from Ajzen (1991) and Johnson et al. (2015).

**Results**

Independent samples t-tests were performed to check the manipulations. All manipulations were successful. Specifically, the high product review score ($M$ = 6.39, $SD$ = .943) was considered significantly higher than the low score ($M$ = 2.03, $SD$ = 1.817) [$t(290.628)$ = -29.598, $p$ < .001]; the large number of reviews ($M$ = 5.20, $SD$ = 1.497) was considered significantly larger than the small number of reviews ($M$ = 3.08, $SD$ = 1.800) [$t(390.989)$ = -12.919, $p$ < .001]; the high involvement product ($M$ = 100.33, $SD$ = 29.434) was considered having a higher involvement level than the low involvement product ($M$ = 71.03, $SD$ = 29.057) [$t(404)$ = -10.093, $p$ < .001].

A three-way ANOVA test was performed. A significant main effect between the product review score and bandwagon perception was discovered $F(1, 398)$ = 99.256, $p$ < .001, partial $\eta^2$ = .200. Additionally, a significant main effect between product involvement and bandwagon perception was unearthed $F(1, 398) = 6.083$, $p = .014$, partial $\eta^2 = .015$. More importantly, a significant interaction effect between the product review score and product involvement for the bandwagon perception was uncovered $F(1, 398) = 7.556$, $p = .006$, partial $\eta^2 = .019$. A post-hoc pairwise comparison with a Bonferroni adjustment was performed to examine the difference between conditions in the interaction effect. The product involvement variable significantly moderated a low review score's influence on the bandwagon perception $F(1, 398) = 13.221$, $p < .001$, partial $\eta^2 = .032$, but the moderating influence of product involvement was not significant when the review score was high $F(1, 398) = .041$, $p = .840$. Specifically, the bandwagon perception was significantly weaker when a low product review score was associated with a high involvement product ($M = 3.30$, $SE = .152$) than when it was associated with a low involvement product ($M = 4.08$, $SE = .151$). The three-way interaction was not significant $F(1,398) = .139$, $p = .709$. Please see Tables 1 and 2 for more information related to the ANOVA test. Please also note that another three-way ANOVA test was performed to examine how the manipulated variables would influence purchase intention. The results were not reported in the body of the manuscript but listed at the end of the manuscript as a note. The purchase intention construct was further analyzed in the PROCESS mediation model, and the associated results were reported in the next two paragraphs.

    The PROCESS model number 4 in the SPSS was employed to analyze the mediation effect. In the model, the bandwagon perception was analyzed as the mediator. The independent variable was the product review score, and the dependent variable was the purchase intention. The results suggest that review scores positively affected the bandwagon perception ($\beta = 1.489$, $p < .001$), and the perception affected purchase intention ($\beta = .816$, $p < .001$). The indirect (i.e., mediation) effect index $= 1.2146$ (95% Bootstrap LLCI $= .9416$ and ULCI $= 1.4980$). Hence, the

bandwagon perception significantly mediated the relationship between the independent and dependent variables (please see Figure 1).

The PROCESS model number 9 in SPSS was utilized to analyze the potential moderation impact of product involvement levels and the number of reviews on the mediation effect. The same mediator, as well as the independent and dependent variables, were examined in the new model. Specifically, the product involvement and number of reviews were examined as two factors that moderated the path between product review scores and the bandwagon perception. The analysis results suggest that the interaction between the product involvement and product review scores positively and significantly affected the bandwagon perception ($\beta = .821$, $p < .01$). The bandwagon perception positively affected purchase intention ($\beta = .816$, $p < .001$). The index revealed a partially significant moderated mediation model with the product involvement as the moderator and the bandwagon perception as the mediator (.6695; 95% Bootstrap LLCI = .1740 and ULCI = 1.1713). Please see Figure 2 for more information about the model.

**Discussion**

The findings of Study 1 confirm that product review scores positively influence the bandwagon perception. A high review score generates a greater bandwagon perception than a low score des. Also, product involvement plays a significant moderating role. A low review score, accompanied by a high involvement product, induces a strong, negative bandwagon perception. The onset of such an effect might be, in part, due to the loss aversion mentality of consumers (Tversky & Kahneman, 1991). When consumers are making high-stakes decisions (e.g., purchasing a high-value product), the potential loss caused by making careless decisions could be consequential. Moreover, the findings implicated the significance of the bandwagon perception as a mediator in affecting the association between product review scores and purchase intention.

**Study 2**

The second study is an extension of the first study. Similar to the first study, one of the goals of Study 2 was to investigate information cues that activate the use of the bandwagon heuristic. A key difference between Study 1 and 2 is the inclusion of the variable, product review valence. In the model examined in Study 2, the bandwagon perception was the mediator, while product review valence was examined as a factor moderating the path between product review scores and the bandwagon perception.

Another key difference between Study 1 and 2 is that the number of reviews was constantly manipulated at a high level across all experimental conditions in Study 2. The decision was made based on the results revealed in Study 1 since the number of reviews was not a significant factor influencing or moderating the impact of review scores on the bandwagon perception. Moreover, a large number of reviews associated with a high or low review score implicates a strong bandwagon support or disapproval of a product. Thus, it would be more meaningful to examine how review valence affects consumers when the potential bandwagon effect is strong than when it is weak. That is, the detrimental influence of negative reviews or the positive influence of positive reviews on consumers is likely to be displayed in a more salient scale when the total number of reviews is large than when it is small.

Furthermore, an additional difference between Study 1 and 2 is the way product involvement was handled in data analysis processes. When analyzing Study 2 data, we utilized product involvement as a grouping variable that divided the proposed moderated mediation model into two parts: the low product involvement model and the high product involvement model. The decision was made for the following reasons. Product involvement should be considered a construct that not only moderates the correlation between review scores and the bandwagon perception but also influences the overall mindset of a consumer and how a consumer processes product-related information. The strength of the moderated mediation effect may vary depending on the level of product involvement. The increase in the level of product involvement heightens the significance of the decision-making outcomes because a high involvement product is usually more important or valuable than a low involvement product, and consumers are likely to be more careful when they evaluate information related to a high-value product than when they evaluate information related to a trivial product. In other words, consumers may spend more mental effort in processing product-related information, and in turn, the information may exert a stronger influence on consumers, when the product involvement level is high than when the level is low. Empirical evidence has proven that information cues are employed by individuals to help them evaluate the information when they perform effortful processing (Chaiken, 1980; Chaiken & Maheswaran, 1994). Some of the empirical findings even suggest that the importance of information cues would enhance as the effort one exerts in processing information increases (Beltramini & Stafford, 1993). Thus, the moderation impact of review valence on the correlation between review scores and the bandwagon perception may vary as the product involvement level changes.

**Method**

A 3 (review valence: all negative vs. half positive and half negative reviews vs. all positive reviews) × 2 (product review scores: low vs. high score) × 2 (product involvement: low vs. high product involvement) between-subjects experiment was conducted on MTurk. The MTurk screening criteria were HIT Approval Rate ≥ 99%, Number of HIT Completed ≥ 1,000,

Location is U.S., and Did not Participate in Study 1 or Pretest. Attention check questions were placed in the questionnaire. The participants who failed to provide a reasonable answer to the attention check questions and those who did not complete the questionnaire were not included in data analysis. Among all participants ($n = 419$), 53.7% were female ($n = 225$) and 45.3% ($n = 190$) were male. The average age of participants was 44, while the average income was $68,786. About 81.6% of participants were White ($n = 342$), 9.1% were African American ($n = 38$), 7.9% were Asian ($n = 33$), and 3.1% were Hispanic/Latino ($n = 13$). Each participant was awarded $1.5 for the participation.

*Stimuli*

The mock website used in Study 1 was utilized and revised. Product reviews were added to the site. The reviews were adapted from real product reviews of bubble bath or laptop sold on online stores such as Amazon, Target, or BestBuy. We investigated the review section of online retailers (e.g., Amazon, Target, BestBuy, etc.). On these online stores, the top page of a product usually contains 5 to 10 reviews, and the reviews are often listed in a chronological manner (i.e., starting from the most recent ones). Thus, we decided to post eight reviews on each of the pages. The wording of the reviews was generic and brief. Each negative review was associated with a 1-star rating score, while each positive review was associated with a 5-star rating score. The usernames of reviewers were generic English names, and the usernames were randomly associated with reviews. The position of reviews on the page was randomly selected. All reviews were labeled as posted "one day ago." Product review scores and product involvement were manipulated the same way as they were in Study 1. Number of reviews was manipulated at a consistent high level (572) for the reasons stated earlier in the method section.

*Procedure of the Experiment*

In the experiment, participants were asked to read a message stating that the purpose of the study was to assist a brand to better market its products online and that fictional brand names were utilized to protect the identity of the brand. Moreover, participants were informed that the reviews were written by actual buyers of the product and that only a selected number of reviews were posted on the site. After reading the messages, each participant was randomly assigned a different webpage URL that would direct each one of them to an experiment condition. Participants were advised to visit the site on their computer for the best viewing experience. Participants were also informed to return to the Qualtrics questionnaire after their visit to the site.

*Measurements*

In addition to the manipulated variables, bandwagon perception ($\alpha = .958$) was measured on the same scale used in Study 1. Purchase intention ($\alpha = .972$) was measured on a 7-point scale (e.g., I would consider buying this product) adapted from Ajzen (1991) and Johnson et al. (2015). Manipulation check questions (e.g., The overall tone of reviews was (1) very negative…(7) very positive) were placed in the questionnaire.

**Result**

Independent samples t-tests and one-way ANOVA test were performed to check the manipulation. All manipulations were successful. Specifically, the high product review score ($M = 5.60$, $SD = 1.847$) was considered significantly higher than the low score ($M = 1.61$, $SD = 1.337$) [$t(372.808) = -25.30$, $p < .001$]. The high involvement product ($M = 89.91$, $SD = 32.78$) was considered having a higher involvement level than the low involvement product ($M = 62.84$, $SD = 30.86$) [$t(417) = -8.703$, $p < .001$]. Negative reviews ($M = 1.73$, $SD = 1.55$) were considered more negative than mixed reviews ($M = 3.45$, $SD = 1.476$) and positive reviews ($M = 6.23$, $SD = 1.552$) [$F(2, 409) = 299.413$, $p < .001$].

A three-way ANOVA test was performed with the bandwagon perception examined as the dependent variable. The product review scores $F(1, 407) = 70.657, p < .001$, partial $\eta^2 = .148$, product involvement $F(1, 407) = 4.071, p = .044$, partial $\eta^2 = .010$, and review valence $F(1, 407) = 106.676, p < .001$, partial $\eta^2 = .344$ were significantly influencing the bandwagon perception. The interaction between the review scores and review valence for the bandwagon perception was also significant $F(2, 407) = 3.019, p = .050$, partial $\eta^2 = .015$. A post-hoc pairwise comparison with a Bonferroni adjustment was performed to further examine the interaction effect. The product review valence significantly and positively moderated the review score's influence on the bandwagon perception in both low score $F(2, 407) = 39.263, p < .001$, partial $\eta^2 = .162$ and high score $F(2, 407) = 70.327, p < .001$, partial $\eta^2 = .257$ conditions. In the low score conditions, we witnessed a steady increase in the bandwagon perception as the review valence transitioned from all negative ($M = 2.12, SE = .161$) to mixed ($M = 3.10, SE = .161$) and to all positive ($M = 4.16, SE = .165$). In the high score conditions, a similar upward trajectory was observed in the bandwagon perception as the review valence transitioned from all negative ($M = 2.96, SE = .181$) to mixed ($M = 4.05, SE = .160$) and to all positive ($M = 5.75, SE = .158$). Other main or interaction effects were not significant. Please see Tables 3 and 4 for more ANOVA test results.

The PROCESS model number 7 in SPSS was utilized to analyze the data. The analyses were divided into two groups based on product involvement levels. When product involvement level was relatively low ($n = 210$), the results revealed that product review scores ($\beta = .863, p < .001$) and product review valence ($\beta = .816, p = .032$) significantly and positively influenced bandwagon perception. Moreover, the bandwagon perception ($\beta = .935, p < .001$) significantly and positively influenced purchase intention. However, the influence of the interaction effect between review valence and review score ($\beta = .273, p = .251$) on the bandwagon perception was not significant. The omnibus moderated mediation effect (see Figure 3) was not significant (.2555; 95% Bootstrap LLCI = -.1990 and ULCI = .7035).

When product involvement level was relatively high ($n = 209$), the results revealed that product review scores significantly and positively influenced the bandwagon perception ($\beta = 1.399$, $p < .001$). Moreover, the bandwagon perception ($\beta = .876$, $p < .001$) significantly and positively influenced purchase intention. Furthermore, review valence significantly and positively moderated the influence of review score on the bandwagon perception ($\beta = .516$, $p = .028$). Finally, the omnibus moderated mediation effect (see Figure 4) was significant with the index = .4520 (95% Bootstrap LLCI = .0290 and ULCI = .8911).

**Discussion**

The findings once again confirm product review scores' significant influence on the bandwagon perception as the increase in review scores improves the overall bandwagon perception of a product. On the other hand, product review valence significantly and positively moderates the association between product review score and bandwagon perception. In the moderated mediation models, which are delineated based on product involvement levels, the review valence significantly and positively moderates the correlation between the review scores and bandwagon perception when the product involvement level is high. Similarly, the proposed moderated mediation effect is significant only when the product involvement level is high.

**General Discussion**

The findings obtained from the experiments reveal intriguing correlations among factors affecting the bandwagon heuristic and purchase intention. The first study focused on examining how product review scores, the number of reviews, and product involvement influence the bandwagon perception. The moderated mediation effects associated with the variables and the purchase intention were also examined. The ANOVA test results confirm the positive influence of product review scores on the bandwagon perception and the significant interaction effect between product review scores and product involvement for the same dependent variable. The findings of the first study also confirm that the bandwagon heuristic is a significant mediator when consumers process information about a product in an online shopping context. The product review score, either by itself or through interaction with the product involvement, affect the dependent variable via the mediation of the bandwagon perception. More importantly, the results of Study 1 reveal that the involvement construct, not the number of reviews, is a significant moderator that affects the correlation between the review score and bandwagon perception. In particular, the increase in the product involvement level amplifies a low review score's negative influence on the bandwagon perception, but the moderating impacts are missing when the review score is high.

Study 2 focused on examining how the review valence, review score, and product involvement affect the bandwagon perception as well as the associated moderated mediation models that have purchase intention as the dependent variable. The ANOVA test results suggest that product review scores positively affect the perceived bandwagon perception, and product review valence positively moderates the influence of the scores on the bandwagon perception. Additionally, the moderated mediation effect that involves the review valence, bandwagon perception, review scores, and purchase intention is significant only when the level of product involvement is high.

Thus, the findings of the two studies support H1. Product review scores positively affect the bandwagon perception. H2 is not supported. The number of reviews does not moderate the impact of review scores on the bandwagon perception. H3 is supported. The bandwagon perception is a significant mediator that affects the correlation between product review scores and purchase intention. H4 is supported. Product involvement moderates product review scores' impact on the bandwagon perception. H5 is supported. Product review valence moderates the correlation between product review scores and the bandwagon perception.

**Theoretical Implications**

Two studies have uncovered significant influences of product review scores, product involvement, and product review valence on the use of the bandwagon heuristic. The findings confirm the conjecture that product review scores are often used by many consumers as an indicator of product quality. The bandwagon heuristic, measured as the bandwagon perception in the studies, has been confirmed as a significant factor that mediates the connection between information cues, such as product review scores, and one's purchase intention of a product.

The findings of the current research can offer insights regarding the mechanism of how the cues (e.g., review score) affect the use of a heuristic and the associated behavioral intention while contributing to future scholars attempting to conduct a meta-analysis on topics related to the bandwagon heuristic or product reviews. Moreover, the research can be regarded as an effort to address the replication crisis faced by many social scientists. For instance, the assertions made by scholars in some studies (e.g., Wang et al., 2023) about the ubiquity of bandwagon cues and bandwagon effects can be confirmed by the findings of the current research.

The more distinctive contribution of the current study lies in its examination of the interaction effects (i.e., moderation and mediation effects). The discovery of the bandwagon heuristic as a significant mediator confirms the assertions made by scholars who have examined heuristics and information cues under the guidance of dual-process models or the MAIN model. That is, heuristic, such as the bandwagon heuristic, is a psychological state employed by individuals when evaluating information (Bellur & Sundar, 2014). As discussed earlier in the literature review, many studies that employed dual process models did not measure the specific heuristic utilized by their study participants (e.g., Eagly & Chaiken, 1975; Wood et al., 1985; Xiao & Myers, 2022). Thus, a major flaw of the reviewed studies is that no direct evidence can be provided to support the notions regarding whether a cue can activate the use of a heuristic and whether the heuristic can help an individual process the information in a certain way. This research study has been constructed partly based on the dual process models. Thus, the measurement and assessment of the bandwagon heuristic as a mediator in the current research can address the limitations identified in previous studies that have employed ELM or HSM.

Though the bandwagon heuristic is often measured in the studies that utilize the MAIN model, the level of involvement, a construct often examined in the dual process models, has not been investigated extensively (e.g., Sundar et al., 2009; Waddell, 2018). Thus, involvement, conceptualized as product involvement and operationalized as two different types of products, has been examined as an important moderator in the current research. The increase in the level of product involvement reflects an increase in product value, relevance, or significance (Petty et al., 1983; Zaichkowsky, 1985). The significant interaction between product involvement and product review score in affecting the bandwagon perception unearthed by the current research is akin to how the usage of heuristic cues affect information processing outcomes during effortful rumination (Chaiken, 1980; Petty & Cacioppo, 1984). The moderated mediation model examined in Study 1 reveals an overall positive interaction between product review scores and product involvement for the bandwagon perception.

Moreover, the findings of the ANOVA results in Study 1 indicate that the interaction between product review scores and product involvement is in a negative direction, and the moderation effect is specifically associated with the low review score scenarios. The findings indicate the ubiquity of the negativity bias phenomenon (Ito et al., 1998; Rozin & Royzman, 2001) as well as the association between the involvement level and the bias because the moderation effect of product involvement is stronger when the product review score is low than when it is high. The negativity bias denotes an individual's tendency to be more attentive to negative events (Rozin & Royzman, 2001). The loss aversion mentality and the rarity of negative events are the potential causes of the bias (Lewicka et al., 1992; Tversky & Kahneman, 1991). Product involvement is inherently connected with the value and usefulness of a product in relation to one's personal life. Price is often an indicator of product value. If a consumer buys an expensive product that receives a low review score, the potential loss or damage associated with the behavior would be greater than if a person buys an inexpensive product with a similar low score. Thus, the loss aversion mentality, possibly stimulated by a low review score, is more likely to affect consumers' judgment when they evaluate a high involvement product than when they evaluate a low involvement product (Tversky & Kahneman, 1991).

    The number of reviews has been examined as another moderator. The findings reveal that the number of reviews does not moderate the influence of product review scores on the bandwagon perception. This finding is contradictory to discoveries unearthed by many existing research articles (e.g., Sundar et al., 2009; Xu, 2013). For example, in the study of Sundar et al. (2009), the bandwagon cue was manipulated as a combination of the number of reviews and review scores. Sundar and colleagues found a significant association between the cue and the bandwagon perception, but the exact degree of impact exerted by each of the factors (i.e., the review scores and review number) was not delineated. The findings of the current study suggest that the bandwagon cues are not equally influential. The number of reviews is not an impactful cue in activating the bandwagon heuristic when the number of reviews is presented in a multi-cue scenario alongside other more impactful factors such as product review scores.

Building upon the findings of Study 1, the second study (i.e., Study 2) analyzed how product review scores, product involvement, and product review valence affect the bandwagon perception. The findings of the ANOVA test indicate that the review valence positively and significantly moderates the influence of review scores on the bandwagon perception. The moderation impact has been further examined in the moderated mediation models. A significant moderated mediation effect has been discovered. The bandwagon perception mediates the correlation between product review scores and purchase intention. The review valence moderates the association between review scores and the bandwagon perception when the level of product involvement is high. The findings demonstrate that the significance and magnitude of the moderated mediation effects vary as the involvement level changes. The increase in product involvement reflects an improvement of product significance. As a result, the decision-making outcomes become more important, driving consumers to be more careful and attentive when evaluating information related to a product. In contrast, the decrease in product involvement weakens the significance of a product. Thus, the decision-making outcomes become trivial, leading consumers to be less attentive to or less caring for the product and its associated information.

When product involvement level is high, the increase in the positivity of reviews can elevate consumers' bandwagon perception. Thus, it can be inferred based on the findings that when a product is important or when it is relevant to a consumer's needs, an increase in review positivity can either salvage a low review score's damage or further enhance a high review score's influence on the image of a product and the associated bandwagon perception. As discussed earlier, product-related information offered in actual reviews is richer and more contextualized than what is offered in numerical review score. Therefore, consumers may regard product reviews either as helpful additions or as more important sources of information than product review scores.

In general, the current research is an in-depth analysis on the cues that activate the use of the bandwagon heuristic in the context of online shopping. Empirical studies in the field of marketing generally discovered an indirect influence of product review scores on purchase intention via a third variable, but the bandwagon perception is usually out of their scope (e.g., Hu et al., 2014; Moe & Trusov, 2011). In contrast, communication-related studies have examined the bandwagon perception as well as its mediation effect, but the examination is often situated in a news consumption context (e.g., Waddell, 2018; Xu, 2013). Thus, a unique contribution of the current study is that the research has connected extant literature in marketing and communication, and the findings of the current research have fulfilled a theoretical gap regarding the moderated mediation effect of how product review scores, with the help of review valence and product involvement, influence purchase intention via the mediation of the bandwagon perception. In addition to theoretical contributions, the findings of the study offer managerial implications that may help brands and marketers better manage their online businesses.

**Managerial Implications**

The findings of the study reveal that a high product review score cues a person to associate the score with the perceived product quality or popularity. Low review scores are proven to be harmful to a person's bandwagon perception and purchase intention. Brands and marketers should strive to maintain positive review scores for the products that they sell online. A few strategies can be employed to improve review scores. First and foremost, brands or companies need to provide high quality products and good service to their customers because a satisfied customer is likely to share the positive experience with others (Hennig-Thurau et al., 2004). Moreover, a brand or marketer should be proactive in soliciting customer feedback. Many companies or marketers send emails to their customers and invite them to leave reviews. Some even offer incentives, such as coupons and discounts, to encourage satisfied customers to post reviews on the retailer's website (e.g., Nike online store) or a third-party site (e.g., Google). Furthermore, brands or marketers can acknowledge customers who have reviewed a product, either by sharing the reviews on their websites or by interacting with reviewers on review sites or social media pages. Doing so may encourage more future customers to leave positive reviews.

Though the number of reviews does not affect a high product review score's influence on the bandwagon perception, the variable significantly affects a low review score's impact. Therefore, brands and marketers must contain the spread of negative information and strive to maintain the positivity of the reviews and overall review scores. One of the strategies to mitigate negative reviews' or low review scores' impact on customers requires brands or marketers to contact the unsatisfied customers in the review section. This would not only allay the dissatisfaction of existing customers but also show potential future customers that the brand cares.

When contacting discontented customers, the brand or company should first sincerely apologize and then offer useful solutions to address the concerns that will eventually resolve the problem. Doing so helps brands or companies retain loyal customers as well as build a positive image as a caring brand. However, brands should also understand how to discern the difference between genuine complaints and malicious attacks initiated by internet trolls or the brand's competitors. The claims made by internet trolls are usually baseless or false. Thus, when dealing with this kind of malicious attacks, brands or companies should use evidence to dispute the false claims. Doing so can prevent other customers, who are unaware of the malicious nature of some negative comments, from spreading false information. As a result, the brands' reputation can be defended.

Moreover, brands or companies that sell high-end products may wish to pay extra attention and effort to maintain a positive brand reputation through product reviews and review scores since a low review score may do more damage to the image of a high-end product than to a low-end product. Perhaps, a brand or company that produces high-value products can hire staff or specialists that will constantly monitor reviews or comments online with the help of information technology (e.g., software). These staff or specialists should be proactive in contacting customers, especially the dissatisfied ones, just as what customer service personnel would do in a retail store.

The current study manipulated eight reviews, which were labeled as recent reviews, and the reviews were presented on a single page. Though only eight reviews were manipulated, the influence of review valence on the correlation between product review score and bandwagon perception was significant when laptop computer (i.e., high involvement product) was presented. The finding indicates that the actual content of review is perhaps more important than mere review score in the eyes of consumers when the product is of high value than when the product is of low value. The finding also suggests that reviews posted on the top page of a product is important. Consumers have limited time or cognitive ability to process information. Thus, many consumers may only read reviews posted on the first page to save time and mental energy in evaluating product quality. Therefore, it is important for marketers and brands to monitor product reviews posted on e-commerce sites especially the ones posted on the first or top page of a product. Doing so would allow marketers or brands to react to adverse situations in a timely manner.

## Limitations and Future Directions

First, experiments were conducted to investigate product review metrics' impact. A future study that connects experimental stimuli with an online database of reviews would better mimic what consumers read on an e-commerce site. In turn, the discoveries would be a more accurate reflection of how online reviews shape consumer impressions. Second, the impact of other metrics, such as review recency, was not investigated in the current study. Future studies may wish to compare the impact of recent reviews against dated reviews on consumers to investigate if review recency would influence the bandwagon perception and purchase intention. Finally, purchase intention, not the actual purchase behavior, was measured. Thus, to better evaluate how product review scores and review valence may affect consumer behaviors, future researchers should connect the questionnaire measurement with e-commerce sales data.

Note:

In Study 1, a three-way ANOVA test was performed to examine the impact of the manipulated variables on purchase intention. A significant main effect between the product review score and bandwagon perception was discovered $F(1, 398) = 46.728$, $p < .001$, partial $\eta^2 = .105$. Furthermore, a significant interaction effect between the product review score and product involvement for the purchase intention was unearthed $F(1, 398) = 4.142$, $p = .042$, partial $\eta^2 = .010$. The three-way interaction and other effects were not significant. The statistics were examined but not reported in the main body of the manuscript because the goal of the examination was on the bandwagon perception.

In Study 2, a three-way ANOVA test was performed to examine the impact of the manipulated independent variables on purchase intention. Significant main effects between the product review score and bandwagon perception was discovered $F(1, 407) = 28.169$, $p < .001$, partial $\eta^2 = .065$, product involvement and bandwagon perception $F(1, 407) = 5.312$, $p = .022$, partial $\eta^2 = .013$, and product review valence and bandwagon perception $F(1, 407) = 54.123$, $p < .001$, partial $\eta^2 = .210$ were uncovered. None of the interaction effects was significant.

## References

Ajzen, I. (1991). The theory of planned behavior. *Organizational Behavior and Human Decision Processes, 50*(2), 179-211.

Atkinson, L., & Rosenthal, S. (2014). Signaling the green sell: The influence of eco-label source, argument specificity, and product involvement on consumer trust. *Journal of Advertising, 43*(1), 33-45.

Bellur, S., & Sundar, S. S. (2014). How can we tell when a heuristic has been used? Design and analysis strategies for capturing the operation of heuristics. *Communication Methods and Measures, 8*(2), 116-137.

Beltramini, R., & Stafford, E. R. (1993). Comprehension and perceived believability of seals of approval information in advertising. *Journal of Advertising, 22*(3), 3-13.

Bode L., Vraga E. K., & Tully M. (2021). Correcting misperceptions about genetically modified food on social media: Examining the impact of experts, social media heuristics, and the gateway belief model. *Science Communication, 43*(2), 225–251

Chaiken, S. (1980). Heuristic versus systematic information processing and the use of source versus message cues in persuasion. *Journal of Personality and Social Psychology, 39*(5), 752–766.

Chaiken, S., & Maheswaran, D. (1994). Heuristic processing can bias systematic processing: Effects of source credibility, argument ambiguity, and task importance on attitude judgment. *Journal of Personality and Social Psychology, 66*(3), 460-473.

Chaiken S., Liberman A., & Eagly A. H. (1989). Heuristic and systematic information processing within and beyond the persuasion context. In Uleman J. S., Bargh J. A. (Eds.), *Unintended thought* (pp. 212–252). Guilford Press.

Sundar, S. S., Xu, Q., & Oeldorf-Hirsch, A. (2009). Authority vs. peer: How interface cues influence users. *Proceedings of the CHI'09 Extended Abstracts on Human Factors in Computing Systems*, 4231-4236.

Sundar S. S., Kim J., & Gambino A. (2017). Using theory of interactive media effects (TIME) to analyze digital advertising. In Rodgers S., Thorson E. (Eds.), *Digital advertising: Theory and research* (pp. 86–109). Routledge.

Sundar, S. S., Molina, M. D., & Cho, E. (2021). Seeing is believing: Is video modality more powerful in spreading fake news via online messaging apps?. *Journal of Computer-Mediated Communication*, *26*(6), 301-319.

Tversky, A., & Kahneman, D. (1991). Loss aversion in riskless choice: A reference-dependent model. *The Quarterly Journal of Economics*, *106*(4), 1039-1061.

Waddell, T. F. (2018). What does the crowd think? How online comments and popularity metrics affect news credibility and issue importance. *New Media & Society, 20*(8), 3068-3083.

Waddell, T. F., & Sundar, S. S. (2017). #thisshowsucks! The overpowering influence of negative social media comments on television viewers. *Journal of Broadcasting & Electronic Media, 61*(2), 393-409.

Wang, S., Chu, T. H., & Huang, G. (2023). Do bandwagon cues affect credibility perceptions? A meta-analysis of the experimental evidence. *Communication Research*, *50*(6), 720-744

Walther J. B., & Jang J. W. (2012). Communication processes in participatory websites. *Journal of Computer‐Mediated Communication*, *18*(1), 2–15.

Westerman D., Spence P. R., &Van Der Heide B. (2012). A social network as information: The effect of system generated reports of connectedness on credibility on Twitter. *Computers in Human Behavior, 28*(1), 199–206.

# Appendix

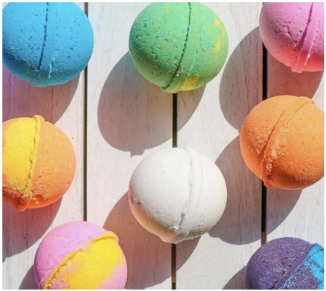

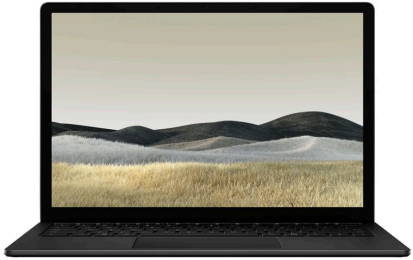

# Future Innovations® Laptop

**1.2** ★☆☆☆☆  4 Votes

**$1200.00**

Quantity
[1]

[Add To Cart]  ♡
[Buy Now]

**Product Details**

- Clean, elegant design
- Case made from durable materials
- Up to 7.5 hours of battery life
- 15.6 inch LED screen

SPECIFICATIONS                +

RETURN & REFUND POLICY        +

SHIPPING INFO                 +

---

STORE            ADDRESS                OPERATING HOURS
Shop All         1845 Belmont Street    Mon - Fri: 10am - 7pm
Shipping & Returns   Andover, MO 67460  Saturday: 10am - 8pm
Store Policy                            Sunday: 11am - 5pm
FAQ

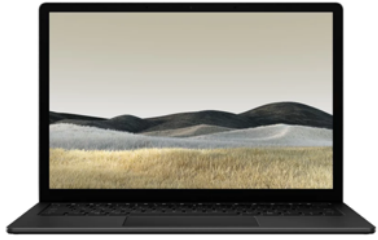

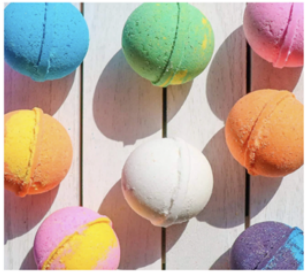

**Figure 1. The Mediation Model in Study 1**

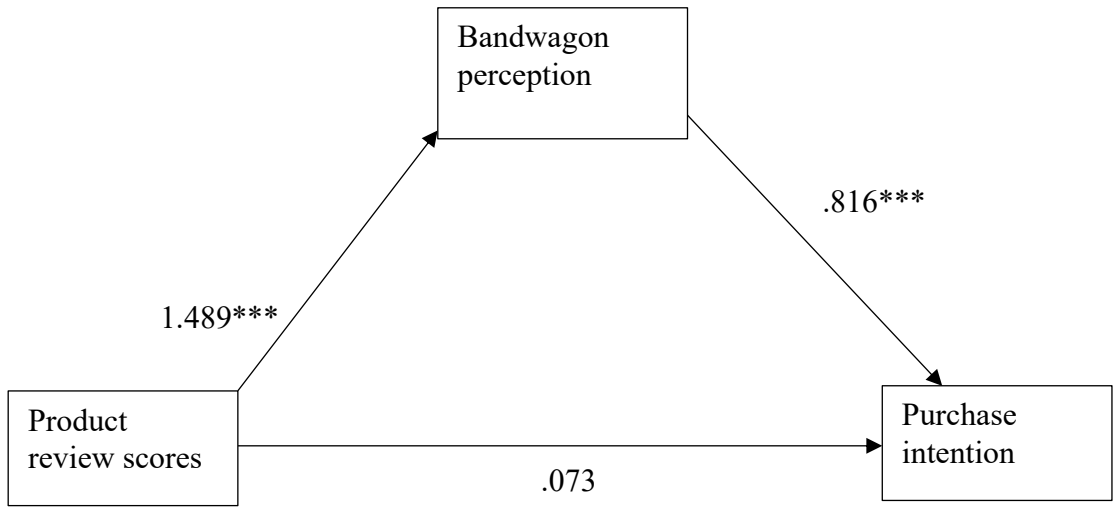

*Note:* mediation index = 1.2146; 95% LLCI = .9416 and ULCI = 1.4980

*\*\*\* p < .001*

**Figure 2. The Moderated Mediation Model in Study 1**

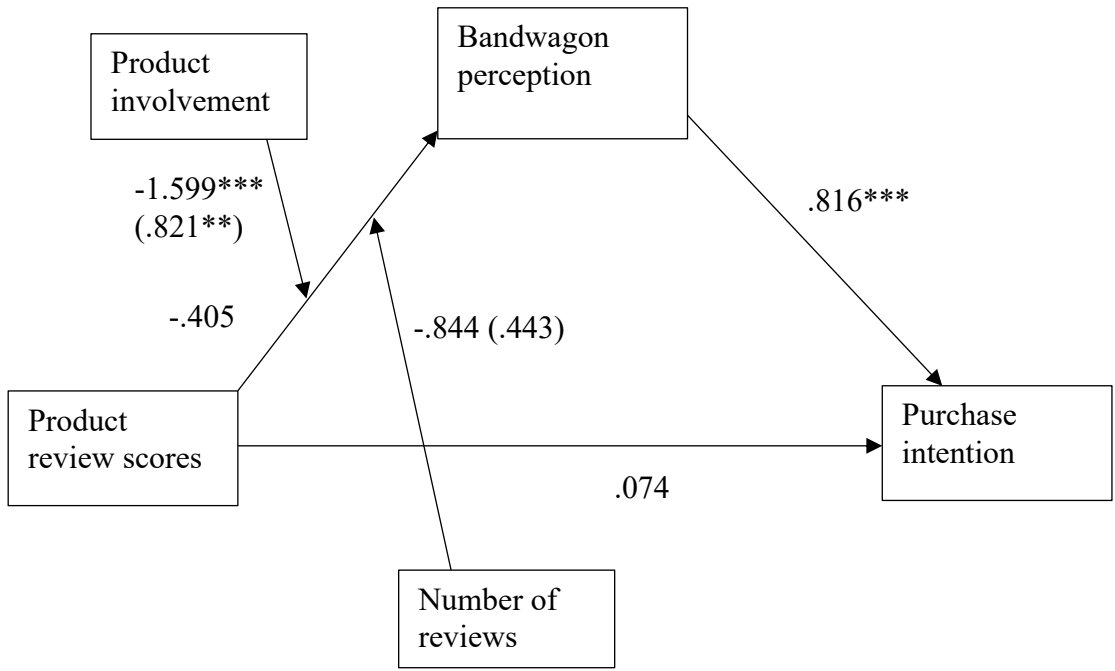

*Note:* moderated mediation index for involvement as the moderator = .6695; 95% LLCI = .1740 and ULCI = 1.1713); interaction effect coefficients between the independent variable and moderators are listed in "()" along the paths; ** *p < .01, *** p < .001*

**Figure 3. The Moderated Mediation Model in Study 2: Low Product Involvement**

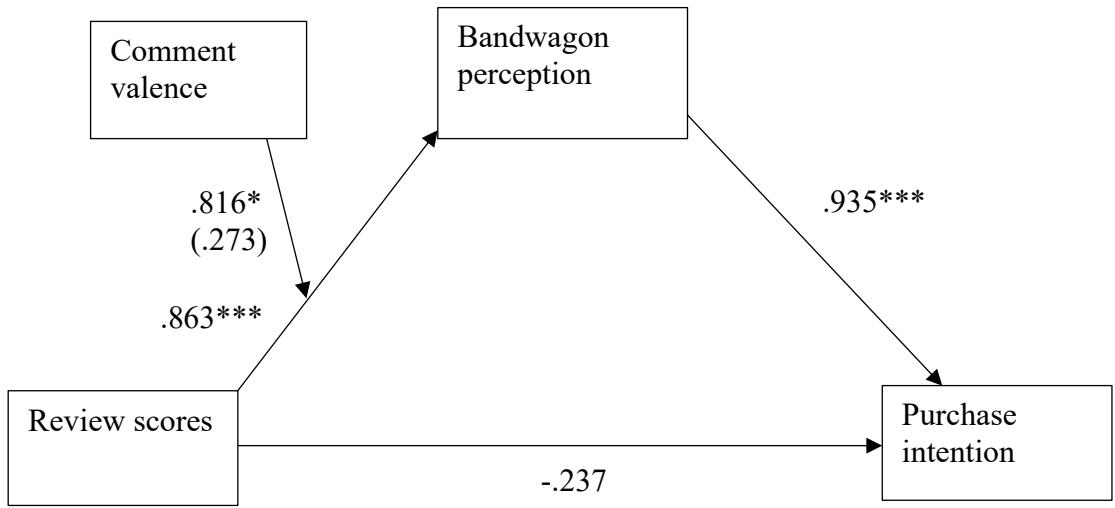

*Note:* moderated mediation index = (.2555; 95% LLCI = -.1990 and ULCI = .7035); interaction effect coefficient between the independent variable and moderator is listed in "()" along the paths

*\* p < .05, \*\*\* p < .001*

**Figure 4. The Moderated Mediation Model in Study 2: High Product Involvement**

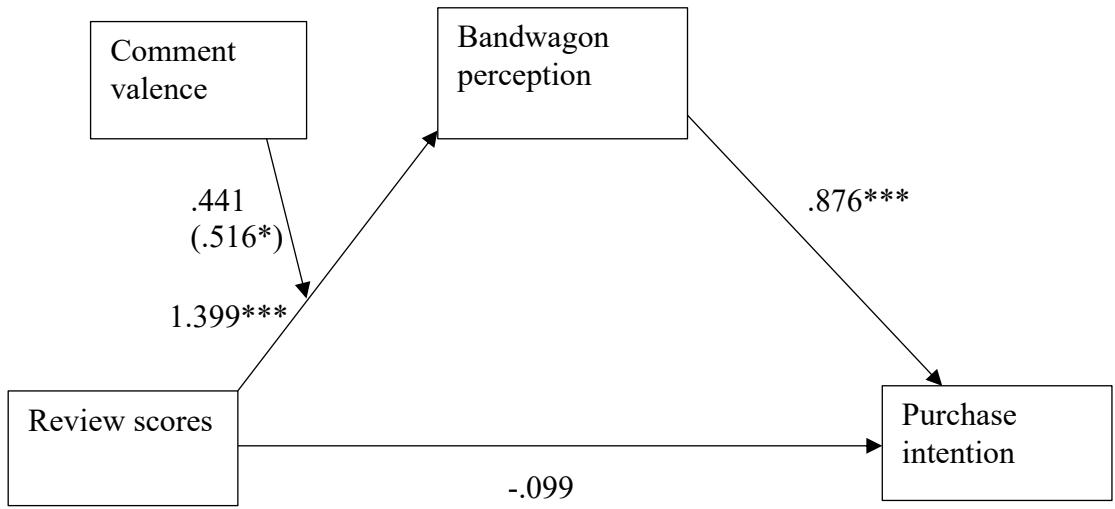

*Note:* moderated mediation index = .4520 (95% LLCI = .0290 and ULCI = .8911); interaction effect coefficient between the independent variable and moderator is listed in "()" along the paths

*\* p < .05, \*\*\* p < .001*

Table 1. Means and Standard Errors of the Three-way ANOVA in Study 1

| | Bandwagon perception | |
|---|---|---|
| | *Means* | *SE* |
| Low product review score | 3.690 | .107 |
| High product review score | 5.180 | .104 |
| | | |
| Small number of reviews | 4.524 | .106 |
| Large number of reviews | 4.346 | .106 |
| | | |
| Low involvement product | 4.619 | .105 |
| High involvement product | 4.250 | .106 |

Table 2. Three-way ANOVA Results of Study 1

| | Bandwagon perception | |
|---|---|---|
| | *F* | *Partial η²* |
| Product review scores | 99.256*** | .200 |
| Number of reviews | 1.419 | .004 |
| Product involvement | 6.084* | .015 |
| Product review scores × number of reviews | 2.197 | .005 |
| Product review scores × product involvement | 7.556** | .019 |
| Product involvement × number of reviews | 1.035 | .003 |
| Product review scores × number of reviews × product involvement | .139 | <.001 |

*p < .05 ** p < .01 ***p< .001

Table 3. Means and Standard Errors of the Three-way ANOVA in Study 2

| | Bandwagon perception | |
|---|---|---|
| | *Means* | *SE* |
| Low product review score | 3.124 | .094 |
| High product review score | 4.254 | .096 |
| | | |
| All negative reviews | 2.538 | .121 |
| Mixed reviews | 3.575 | .114 |
| All positive reviews | 4.955 | .114 |
| | | |
| Low involvement product | 3.825 | .094 |
| High involvement product | 3.554 | .096 |

Table 4. Three-way ANOVA Results of Study 2

| | Bandwagon perception | |
|---|---|---|
| | *F* | *Partial η²* |
| Product review scores | 70.657*** | .148 |
| Product review valence | 106.675*** | .344 |
| Product involvement | 4.071* | .010 |
| Product review scores × review valence | 3.019* | .015 |
| Product review scores × product involvement | 3.818 | .009 |
| Product involvement × review valence | 1.077 | .005 |
| Product review scores × review valence × product involvement | .227 | .001 |

*p < .05 ***p< .001